# Long-distance spin transport in frustrated hyperkagome magnet Gd$_3$Ga$_5$O$_{12}$

Di Chen[1], Bingcheng Luo[2], Lei Xu[3], Zian Xia[4], Linhao Jia[2,1], Shaomian Qi[2], Congkuan Tian[2,1], Kangyao Chen[2], Hang Cui[2], Guangyi Chen[2], Shili Yan[1], Miaoling Huang[1], Jian Cui[1], Ya Feng[1], Zhentao Wang[3,5]*, Jiang Xiao[4,6]*, Jianhua Zhang[7,8], Ryuichi Shindou[2], X.C. Xie[2,6,9], Jian-Hao Chen[2,1,9,10]*

[1]Beijing Academy of Quantum Information Sciences; Beijing, 100193, China.

[2]International Center of Quantum Materials, School of Physics, Peking University; Beijing, 100871, China.

[3]School of Physics, Zhejiang University; Hangzhou, 310058, China.

[4]Department of Physics and State Key Laboratory of Surface Physics, Fudan University; Shanghai, 200433, China.

[5]Center for Correlated Matter, Zhejiang University; Hangzhou, 310058, China.

[6]Institute for Nanoelectronic Devices and Quantum Computing, Fudan University; Shanghai, 200433, China.

[7]N.O.D topia Biotechnology Co., Ltd.; Guangzhou, 510000, China

[8]Simpcare Biotechnology Co., Ltd.; Guangzhou, 510000, China

[9]Hefei National Laboratory; Hefei, 230088, China.

[10]Key Laboratory for the Physics and Chemistry of Nanodevices, Peking University; Beijing, 100871, China.

*Corresponding authors: Zhentao Wang (ztwang@zju.edu.cn); Jiang Xiao (xiaojiang@fudan.edu.cn); Jian-Hao Chen (chenjianhao@pku.edu.cn)

## Abstract

Transport of spin angular momentum over large distance has been a long sought-after goal in the field of spintronics. While the majority of the research effort has been devoted to the spin transport properties of magnetically ordered materials, spin transport in magnetically frustrated materials has received little attention. Here, we report an anomalous state in frustrated hyperkagome magnetic insulator Gd$_3$Ga$_5$O$_{12}$, where spin angular momenta can be transported over a long distance of 480μm, far exceeding the transport distance of any diffusive spin current in magnetically ordered materials, to the best of our knowledge. Monte Carlo simulations reveal significant spin fluctuations, spin-spin correlations and an absence of conventional magnons in such anomalous state; while the response of the anomalous state to perturbation is found to be akin to an overdamped forced oscillator. We find close relation of such state to the correlated "director" state in the material. Our result provides an effective electrical technique to characterize spin-spin correlations and frustrations; it also unveils the potential of frustrated magnets as powerful channel materials for spin transport.

# Introduction

The transport of spin angular momenta in solids constitutes the heart of spintronics technology, a branch of electronics that utilizes the spin degree of freedom for information processing [1,2,3]. In particular, magnons in insulating ferromagnets and antiferromagnets have been identified as an efficient carrier for spin current [4,5,6,7,8], opening a new field of magnon-based spintronics. More recently, spin transport was observed in paramagnetic insulators [9,10,11] and antiferromagnets above the Néel temperature [12,13,14], suggesting that long-range magnetic order may not be a prerequisite for spin transport. Unlike ordinary magnetically ordered materials, frustrated magnets, an important class of magnetic materials where long-range magnetic orders could be prevented due to geometric [15] or exchange [16] frustrations, are largely unexplored in the thriving field of spintronics [17,18].

Frustrated systems without magnetic order do not have well-defined magnons, but instead have intrinsic spin fluctuations that support novel magnetic states, such as spin ice, spin liquid, and other exotic quantum states (as schematically illustrated in Fig. 1a) [19]. $Gd_3Ga_5O_{12}$ (GGG) is a typical example of frustrated magnetic insulators. High quality single crystal GGG is widely available since it is a common substrate material [20] for growing epitaxial garnet thin films like yttrium iron garnet (YIG). At room temperature, GGG behaves as an ordinary paramagnetic insulator and does not affect either the charge or spin transport in any magnetically ordered materials that are grown on it. At low temperatures, GGG exhibits characteristics of a geometrically frustrated antiferromagnetic insulator, which presents strong spin fluctuations without static order down to 25mK [21]. It exhibits complex magnetic phases at low temperatures, where different candidates including spin glass [22], spin liquid [23] and a "hidden order" of ten-spin-loop director state [24,25] are proposed. As shown in Fig. 1b, this "hidden order" state is composed of loops of ten individually fluctuating magnetic $Gd^{3+}$ ions, where a long-range "director state" with no overall magnetic moment can be defined by the remanent anti-ferromagnetic coupling among the disordered spins [24]. Such state does not break any crystal structural symmetry or time reversal symmetry, yet its correlations could exist to moderate magnetic fields [26,27].

In this study, we experimentally investigate the phase diagrams of the nonlocal second harmonic spin transport (SHST) signal in GGG as a function of injection current, magnetic field and temperature. In addition to a normal SHST signal similar to the magnon transport signal in magnetically ordered systems, an anomalous spin signal is detected below 5K and for magnetic field ≲ 9T, where the spin fluctuation and spin-spin correlation are strong [24, 28]. This anomalous spin transport state is characterized by a frustration induced correlation process and exhibits a long spin transport distance of up to 480μm, far exceeding the transport distance of any diffusive spin current in magnetically ordered materials [5, 6]. Monte Carlo simulations reveal significant spin fluctuation and spin-spin correlation under conditions when long-distance spin transport emerges experimentally, and the low energy spin excitations response to electrical perturbation in a way that is akin to an overdamped forced oscillator. Our experimental result unveils the potential of frustrated magnets as promising candidates for spin transport.

## Results

Fig. 1c displays a simplified atomic schematic of the hyperkagome structure of GGG, where only the magnetic $Gd^{3+}$ ions are shown for clarity. These $Gd^{3+}$ ions form two interpenetrating lattices of corner-sharing triangles [29], providing the basic ingredient of geometric frustration. The absence of long-range magnetic order down to a temperature of 2K [21], and a Curie-Weiss temperature of -2.05K [10, 30] of the GGG <111> sample is experimentally verified by magnetic moment versus temperature measurement (Supplementary Information Fig. S1a). The field-dependent in-plane magnetization gives a low-temperature saturation magnetization of ~7μB per $Gd^{3+}$ (Supplementary Information Fig. S1b), in agreement with previous studies [11]. A comparison of the magnetization data at $T \leq 2K$ to the Brillouin function for non-interacting and isotropic S=7/2 spins reveals a deviation (Supplementary Information Fig. S1c). This deviation from pure paramagnetic behavior becomes more pronounced at lower temperatures, suggesting the dominance of spin-spin frustration and correlations in GGG [31].

**Nonlocal spin transport phase diagrams**

Fig. 1d presents a schematic diagram of the experimental setup used in this study. The spin transport device comprises of two platinum (Pt) wires, an injector, and a detector, which are fabricated on a GGG <111> channel. An AC current $I_{in} = I_{in,0} sin\omega t$ with frequency $\omega$ is applied through the injector, creating thermal spin excitations in the channel. Here $I_{in,0}$ is the magnitude of the AC injection current. The nonlocal second harmonic inverse spin Hall voltage $V_{2\omega}$ is measured by lock-in amplifiers from the detector [5, 32]. To comply with the orthogonal rule of the inverse spin Hall effect [32], an in-plane magnetic field with varying angle $\theta$ is applied to produce an in-plane **M** component, and a $cos\theta$ dependence of $V_{2\omega}$ can be detected as shown in the inset of

Fig. 1d. GGG has an energy bandgap of approximately 5.66 eV [33], which excludes any contribution of conduction electrons in the spin signal. We also exclude the current leakage effect by measuring the resistance between two Pt electrodes in the GGG devices in Supplementary Information Fig. S2. When measuring the AC $V_{2\omega}$ signal, the in-phase component $V_{2\omega}^X = |V_{2\omega}|\cos\phi$ and the out-of-phase component $V_{2\omega}^Y = |V_{2\omega}|\sin\phi$ are obtained from the lock-in amplifier, where $\phi$ is the relative phase angle of this AC signal to the injection AC current. The sign of the SHST signal $V_{2\omega}$ is defined by the sign of $\sin\phi$, i.e., $V_{2\omega} = sign(\sin\phi)|V_{2\omega}|$.

Figs. 2a and 2b show the mapping of $V_{2\omega}^X$ and $V_{2\omega}^Y$ measured at $T = 2K$ as a function of magnetic field **B** and injection current $I_{in}$, respectively. The magnetic field ranges from 0T to 14T, and the injection current ranges from 0μA to 132μA. Here, the detector is positioned 2μm away from the injector and the in-plane magnetic field **B** is directed at an angle of $\theta = 0$, where $\theta$ represents the angle between **B** and the perpendicular direction of the detector Pt electrode (as depicted in Fig. 1d). For SHST signals in conventional magnetic systems, $V_{2\omega}^X = 0$ and $V_{2\omega}^Y > 0$, corresponding to $V_{2\omega} > 0$ and $\phi = \frac{\pi}{2}$. This value of $\phi$ is due to the second harmonic nature of the signal, i.e. $V_{2\omega} \propto -(I_{in,0}\sin\omega t)^2 \propto \frac{I_{in,0}^2}{2}\sin(2\omega t + \phi)$, where $\phi = \frac{\pi}{2}$. The negative sign in the above formula comes from the convention in the field to present thermal magnon signals as positive signals at $\theta = 0$ [5].

Thus, the region in Figs. 2a and 2b where $V_{2\omega}^X = 0$ and $V_{2\omega}^Y > 0$ represents the normal state of GGG, of which the SHST response is similar to conventional magnetic systems; while the region with $V_{2\omega}^X \neq 0$ or $V_{2\omega}^Y < 0$ marked the emergence of an anomalous state in the spin transport of the material. Since the normal and anomalous $V_{2\omega}^Y$ are of opposite signs and could compete with each other while they coexist, the anomalous state of GGG is best characterized by a non-zero $V_{2\omega}^X$. Control experiments with Cu electrodes are shown in Supplementary Information Fig. S3, which exclude any possible reactive components contributed to $V_{2\omega}^X$. We define the anomalous phase as the region in Fig. 2a below the gray dashed line. This line marks a cut-off value of 10% of the maximum $V_{2\omega}^X$ signal. Note that other definitions of cut-off value of non-zero $V_{2\omega}^X$ yield similar results.

Fig. 2b shows $V_{2\omega}^Y$ versus **B** and $I_{in}$ with the same gray dashed line as in Fig. 2a. Above the gray line, $V_{2\omega}^X = 0$ and $V_{2\omega}^Y > 0$, representing the normal state of GGG. Below the gray line, there are two areas: one with $V_{2\omega}^X \neq 0$ and $V_{2\omega}^Y > 0$ at smaller $I_{in}$; the other with $V_{2\omega}^X \neq 0$ and $V_{2\omega}^Y < 0$ at larger $I_{in}$. These two areas belong to the same region where the anomalous state coexists with the normal state with different relative strengths. Equivalently, one can also plot $V_{2\omega}$ versus **B** and $I_{in}$ which can be found in Supplementary Information Fig. S4a. The magnetic field angle $\theta$ dependent $V_{2\omega}$ and $\phi$ of the typical normal and anomalous states can be found in Supplementary Information Fig. S4b-e, highlighting the stark differences between the two states. Data from another GGG crystal can be found in Supplementary Information Fig. S5, further verifying the existence of the anomalous state. Note the signal $V_{2\omega}$ vanishes at $\theta = 90°$ and $270°$ in all the range of magnetic field we studied (Supplemental Figure S5-1), which is consistent with the inverse spin Hall effect rather than spurious reactive components of the circuit.

In Figs. 2c and 2d, $V_{2\omega}^X$ and $V_{2\omega}^Y$ versus **B** with the magnetic field angle $\theta = 0$ and $T = 2K$ are plotted for GGG and MnPS$_3$, respectively. MnPS$_3$ is a conventional magnetic insulator with antiferromagnetic order and it has a normal SHST signal resembling that of YIG [34]. As shown in Fig. 2c, $V_{2\omega}^X > 0$ in GGG as soon as **B** > 0; it increases with increasing **B**, reaching a maximum, then decreases and finally diminishes above ~9T. In contrast, $V_{2\omega}^X$ in MnPS$_3$ remains zero from 0T to up to 14T, as expected for conventional magnets [34]. For the $V_{2\omega}^Y$ component, $V_{2\omega}^Y < 0$ in GGG as soon as **B** > 0; it decreases with increasing **B**, reaching a minimum, then increases and reverses sign to become positive above ~9T. In comparison, $V_{2\omega}^Y$ in MnPS$_3$ is positive, increases with increasing magnetic field, then decreases slightly, but remains positive up to 14T as expected. One can find that the $V_{2\omega}^{X,Y}$ vs. $B$ of GGG from 9T to 14T is similar to that of MnPS$_3$ from 0T to 14T (normal state), while the $V_{2\omega}^{X,Y}$ vs. $B$ curves of GGG below 9T are completely different from the normal behavior, indicating an anomalous state in the material. Note that a negative $V_{2\omega}^Y$ in GGG means that the spin transport direction is opposite to those in conventional magnetic materials when other excitation conditions are the same, highlighting the prominent feature of the anomalous state.

Figs. 2e-2f exhibit $V_{2\omega}^X$ and $V_{2\omega}^Y$ versus $I_{in}$ with $\theta = 0$ and $T = 2K$, for GGG and MnPS$_3$, respectively. For GGG, finite $V_{2\omega}^X$ emerges with minimal $I_{in}$ and monotonically increases with increasing $I_{in}$, indicating that the anomalous state is present irrespective of the injection current; $V_{2\omega}^Y$ is positive only in the low current range, then quickly decreases to zero and changes sign, finally saturating at large $I_{in}$, exemplifying the competition of the normal and anomalous states in the $V_{2\omega}^Y$ component. In contrast, for MnPS$_3$, $V_{2\omega}^X$ is always zero, and $V_{2\omega}^Y > 0$ up to a large

injection current. The deviation from the quadratic relationship at larger $I_{in}$ could be attributed to non-linear spin Seebeck effect [34].

The next intriguing question is how this anomalous state behaves at different temperature. Fig. 3a and Fig. 3b show the 2D plots of $V_{2\omega}^X$ and $V_{2\omega}^Y$ vs. $T$ and **B** with $\theta = 0$, $I_{in} = 100\ \mu A$ and a channel length of $1\mu m$ (see Supplementary Information Fig. S7 for data with different channel lengths). While the normal SHST signal (red part in Fig. 3b) persists up to 14K, the anomalous SHST signal (blue parts in Figs. 3a-b) vanishes above ~5 K and above ~9 T. Note that the nonzero $V_{2\omega}^X$ region in Fig. 3a does not exactly coincide with the negative $V_{2\omega}^Y$ region in Fig. 3b, because of competition between the negative (anomalous) signal and the positive (normal) signal in the $V_{2\omega}^Y$ channel, similar to the situation in Fig. 2b. To demonstrate the anomaly, $V_{2\omega}^X$ and $V_{2\omega}^Y$ versus $T$ for GGG at $B$ = 3T are shown in Fig. 3c; similar data taken from MnPS$_3$ are shown in Fig. 3d. For MnPS$_3$, $V_{2\omega}^X$ is strictly zero in the whole temperature range; positive $V_{2\omega}^Y$ emerges with decreasing temperature below 20K and continue to increase [34], which is expected for conventional SHST signal. For GGG, however, finite $V_{2\omega}^X$ emerges at about 7K and continue to increase with decreasing temperature; positive $V_{2\omega}^Y$ emerges at about 12K and initially increases with decreasing temperature (similar to MnPS$_3$); a strong down turn in $V_{2\omega}^Y$ appears at low temperature, creating a maximum in the signal and causing the signal to reverse sign at 3K. The above data clearly indicates that anomalous spin transport state dominates at lower magnetic fields and at lower temperatures.

**Correlated spin dynamics in the anomalous state**

From the previous section, two distinct spin transport states were found in GGG, i.e., one normal state and one anomalous state. The normal state is characterized by: 1) $V_{2\omega}^Y > 0$ and $V_{2\omega}^X = 0$; 2) signals persist up to over 10K and up to 14T. The GGG channel right below the injector is lightly perturbed thermally, the temperature oscillation $T(t) = T_0 + \Delta T sin(2\omega t + \frac{\pi}{2})$ changes the chemical potential of the magnons below the injector. A chemical potential difference $\Delta\mu$ is established in the spin transport channel, and a spin current $j_s = G\Delta\mu(T)$ is generated, where $G$ is the magnon conductance. For the normal paramagnetic phase above 5K, the magnetic moments are not entangled with each other and are free to relax, so that the system responds effectively instantaneously, producing no measurable $V_{2\omega}^X$ signal. In this regime, since the nearest-neighbor exchange interaction in GGG is small with $J \sim 0.1$ K [35], dipolar interactions become significant and could support spin-wave-like excitations under magnetic field, attributing to the normal signal in GGG [11].

In contrast, the anomalous state manifests itself with distinctive SHST signals: 1) $V_{2\omega}^Y < 0$ and $V_{2\omega}^X \neq 0$; 2) signals emerge below 5K and below a magnetic field of ~9T. The emergence of this anomalous state at low temperature points its origin to the geometric frustrations in GGG, and exclude the possibility of phonon-carried spin current which should sustain to high temperatures [36, 37]. Similarly, the suppression of the anomalous state at and above saturating magnetic fields is natural as geometric frustrations become irrelevant in the presence of strong Zeeman energy [11, 38]. What's more, in conventional magnets, thermal injection weakens the magnetization near the injector, resulting in the transport of spin angular momenta in opposite direction than the expectation for an electrically injected spin current. Since so far thermal magnon induced spin transport is studied almost solely in conventional magnets, a convention has been established that such opposite direction of spin transport would provide $V_{2\omega}^Y > 0$ [5]. Note that careful consideration has to be given for the thin film systems considering the accumulation of spin excitations at the boundaries [39], but for the bulk system studied here the presented sign convention is applicable. For the anomalous state in our work, a moderate magnetic field could not align the frustrated magnetic moments and still results complicated spin configurations with strong spin-spin correlations. Thermal perturbation in frustrated magnets will weaken the intrinsic frustration and spin-spin correlation effects, resulting in more magnetization along the magnetic field direction. Thus, the anomalous state may transport angular momenta in the same direction as the applied field, resulting in $V_{2\omega}^Y < 0$ according to the above-mentioned convention, suggesting a crucial role of spin fluctuations. Indeed, the only work in the literature that is related to correlated spin transport (one-dimensional spinon local spin Seebeck effect), also shows $V_{2\omega}^Y < 0$ [40].

The finite $V_{2\omega}^X$ in the anomalous state indicates an additional phase delay $\Delta\phi$ in $V_{2\omega} \propto j_s \propto \sin(2\omega t + \frac{\pi}{2} + \Delta\phi)$, which is related to a retardation in magnetic relaxation induced by spin-spin correlations. In a correlated spin system, flipping one spin requires flipping many other spins in order for the system to go from one ground state to another, resulting in the phase delay analogous to the imaginary part of AC susceptibility [41]. As shown in Fig. 4c, Monte Carlo simulation of magnetic susceptibility also indicates strong spin fluctuations as well as spin-spin correlations below 5K and near the saturation field, which corresponds well to the $V_{2\omega}^X$ distribution shown in Fig. 3a. Fig. 4a shows $V_{2\omega}^X$ and $V_{2\omega}^Y$ versus injection current frequency $\omega$ with in-plane magnetic field at angle $\theta = 0$ and magnitude $B = 3T$ at temperatures in the range of 1.8 K - 4 K. A maximum in $V_{2\omega}^X$ appears in the $V_{2\omega}^X$ vs. $\omega$ curve at low temperatures, representing a quasi-resonance condition of the frustrated and correlated spin system under external excitation. Moreover, such maximum in $V_{2\omega}^X$ moves from 37Hz to 60Hz and declines in magnitude with increasing temperature and eventually disappears at 4K, which is in line with the fact that increasing temperature weakens the effects of spin correlations, resulting in an increasing quasi-resonance frequency and an overall decreasing signal from spin-spin correlations. Indeed, geometric frustrations in GGG lead to macroscopically multi-degenerate spin configurations with mixed total magnetization [27]. One effect of the thermal injection is changing the spin configurations; with spin-spin correlations, the spin configurations (thus magnon chemical potential) could not always keep up with the change of the temperature and is delayed by $\Delta\phi$, where the phase delay is relevant to $\omega$. Such a response of the chemical potential to the temperature modulation is mathematically equivalent to an overdamped forced oscillator:

$$\ddot{x}(t) + \beta\omega_0\dot{x}(t) + \omega_0^2 x(t) = \delta T^0 e^{-i\omega t},$$

where the chemical potential difference $\Delta\mu$ is related to the speed of response $\dot{x}$ as $\Delta\mu \propto \dot{x}/\omega_0$, and the eigen frequency $\omega_0(T, B)$ is related to the evolution of spin configuration depending on both temperature and the external magnetic field. Then the response function in the frequency domain can be written as:

$$\chi(\omega) = \frac{i\omega/\omega_0}{\omega_0^2 - \omega^2 + i\beta\omega_0\omega}.$$

Here $\chi(\omega) \propto \Delta\mu$, and the real and imaginary parts of $\chi(\omega)$ are related to $V_{2\omega}^Y$ and $V_{2\omega}^X$, respectively, with a corresponding phase delay $\phi = arccot\frac{\beta\omega_0\omega}{\omega_0^2 - \omega^2}$. The lower the temperature and the stronger the correlation would result in lower magnetic relaxation eigen frequency $\omega_0$, as the magnetic moments are increasingly affected by the frustrations and behave cooperatively in clusters [25]. Fig. 4b displays the simulated $V_{2\omega}^Y$ and $V_{2\omega}^X$ with different $\omega_0$, which agrees well with the experimental results shown in Fig. 4a. This indicates that the overdamped forced oscillator model captures the main physics of the correlated spin response in GGG under excitation.

**Long-distance spin transport in GGG**

One particularly intriguing question is how spin-spin fluctuation and correlation in frustrated magnetic insulators affect the spin transport properties of such materials, which has never been probed experimentally before. Fig. 5a plots the distance $d$ dependent $|V_{2\omega}|$ in GGG at one representative point in the anomalous state ($T = 2K, B = 3T, I_{in} = 200\mu A$) and two representative points in the normal state (strong magnetic field: $T = 2K, B = 14T, I_{in} = 100\mu A$ and high temperature: $T = 5K, B = 3T, I_{in} = 100\mu A$). Fig. 5a also plots a typical distance dependent $|V_{2\omega}|$ in MnPS$_3$, which is a conventional magnetically ordered insulator. The inset in Fig. 5a is a zoom-in plot of all the data except for the anomalous state. It can be seen that the normal state GGG has a spin diffusion length of several micrometers. For example, at $T = 5K, B = 3T, I_{in} = 100\mu A$, we extract a spin diffusion length $\lambda = 1.7087 \pm 0.235\ \mu m$ via a 2D diffusion model (details in Methods and Supplementary Information Fig. S9), which is comparable to the previous study [11] in GGG at $T = 5K, B = 3.5T$ with $\lambda = 1.8 \pm 0.2\mu m$ obtained via the same model. However, the anomalous SHST signal in GGG decays considerably slower and remains finite up to a distance of $480\mu m$, two orders of magnitude larger than that of the normal state GGG and MnPS$_3$. An exceptionally long spin diffusion length $\lambda = 209.19 \pm 22.6\mu m$ is extracted from the 2D diffusion model (see Supplementary Information Fig. S10).

With the analysis in the previous section, frustration induced novel spin states could be the cause of the anomalous signal in GGG. Since our experimental temperature is well above the temperatures for spin glass [22] and spin liquid [23] transitions in GGG, the most likely origin of the anomalous state should be rooted in the hidden-order state reported in GGG and its isostructural crystal GAG (Gd$_3$Al$_5$O$_{12}$) [24, 28]. This hidden-order composed of ten Gd$^{3+}$ ions arranged in a ring shape, which forms a nematic director state under the antiferromagnetic interactions and local *x-y* anisotropic interactions [24, 25]. Neutron scattering revealed the onset of director correlations in GGG/GAG below 5K [28, 42], which agrees with the onset of the anomalous spin transport in our experiment. More importantly, the ten-spin directors state could give rise to long-range correlations as shown in Fig. 1b, i.e., the multipoles (indicated by orange double-headed arrows) located at the center of the ten Gd$^{3+}$ ions rings are highly correlated, which could mediate long-distance spin transport. It has been theoretically and experimentally shown that the ten-spin directors state correlations still play a role in the material under a moderate magnetic field [26, 43]. Fig. 4d shows the magnon life time of the highest band at Γ point as a function of magnetic field obtained from our Monte Carlo simulation. The magnon life time decreases with decreasing field and a crossover happened from the paramagnetic state to the frustrated state with spin-spin correlations (details in Methods and Supplementary Information Fig. S13-14). Note that non-spin wave excitations with lower excitation energy could exist at lower field, since multi-degenerate local minima are found in the classical spin configurations [27]. Such low energy excitations with correlations might be the cause of the long-distance spin transport in GGG. Interestingly, AC susceptibility measurements on GGG have shown two distinct thermally activated processes, one related to single spin reversal and the other to the fluctuations of ten-ion loops [28], which may correspond to the normal and anomalous spin transport states we observed.

## Discussion

Lastly, we emphasize in Fig. 5b the exceptional spin transport capability in the anomalous state in GGG, as compared to thermally injected spin current in other magnetic materials [5, 6, 44, 45] with similar injection power (see Supplementary Information Fig. S16). These materials include antiferromagnets and ferrimagnets such as $\alpha$-$Fe_2O_3$ [6] as well as YIG [5]. It is interesting to see that frustrated magnet GGG without magnetic order can transport spin angular momenta to much greater distance than any magnetically ordered materials. Actually, long-range spin transport in disordered magnetic systems has been theoretically studied [46, 47, 48], and the ingredient of antiferromagnetic interactions and local *x-y* anisotropic interactions has led to the prediction of superfluid transport in frustrated magnets [49]. Experimentally, we found the signal decaying at a slower rate than the fitting curves as shown in Fig. 5a and in Supplementary Information Fig. S10, pointing to a different dissipation mechanism from conventional spin diffusion in anomalous state GGG, and revealing the great potential for the innovative mechanism of spin transport through spin fluctuations and correlations.

In conclusion, we experimentally studied the nonlocal spin transport in frustrated hyperkagome magnetic insulator GGG where magnetic order is prevented by geometric frustrations. We report an anomalous spin transport state in GGG at low temperature, with ultra-long-distance spin transport of over 480μm. Such low-dissipation transport distance is significantly longer than that in the normal state GGG and those achieved in any known magnetically ordered materials. Theoretical simulations indicate close connections between the anomalous spin transport state with the spin fluctuation and spin-spin correlation in GGG. Our result provides a valuable tool for studying spin fluctuations in frustrated magnets, elucidates the immense potential of frustrated magnets as promising materials for spin transport channels, and potentially enables novel spintronics devices.

# Methods

**Device fabrication and magnetization characterization**

Gd$_3$Ga$_5$O$_{12}$ <111> single crystals slabs with thickness of 500μm were commercially purchased from Hefei Kejing Materials Technology Co., Ltd. and CRYSTAL GmbH. The injector and detector electrodes in the nonlocal spin transport devices are fabricated with standard electron-beam lithography, platinum deposition and lift-off processes. Platinum is deposited in a magnetron sputtering system, and the length of the electrodes is 100μm, the width of the wires is ~200nm with a thickness of 10nm. Afterwards, 5nm of titanium and 50nm of gold are patterned to contact the platinum wires. Distances between the platinum wires range from 1μm to 480μm center-to-center. Data from nonlocal spin transport GGG devices fabricated on five GGG slabs (device#1 – device#5) were presented. Data shown in Fig. 2a-b, Fig. 3a-b were obtained from device#1; data shown in Fig. 2c, 2d, 3e and the normal signal taken at 2K shown in Fig. 5a were obtained from device#2; the normal signal taken at 5K shown in Fig. 5a were obtained from device#3; data shown in Fig. 4a were obtained from device #4; the anomalous signal shown in Fig. 5a were obtained from device#5. Others are stated in the Supplementary Information. For magnetization measurements, the GGG slab cut into 2.5mm long and 2mm wide was measured using SQUID magnetometry in a temperature range from 0.4 K to 300 K under external magnetic fields up to 7T.

**Nonlocal spin transport measurement**

The spin transport measurement of GGG is done in PPMS and Oxford TeslatronPT with low-frequency lock-in amplifier technique. The injection AC current in the range from $0\mu A$ to $500\mu A$ is provided by lock-in amplifier (Stanford Research SR830). Except for frequency dependent measurements (0.77-200.77Hz), the injection current frequency is 17.77Hz for all the experiments. Also, lock-in amplifiers (NF LI5650 and Stanford Research SR830) are used to probe the nonlocal voltages amplified by low noise voltage preamplifiers (NF LI75A). In lock-in AC measurements, four components of the $V_{2\omega}$ can be obtained, including the amplitude $R$ component $V_{2\omega}^R = |V_{2\omega}|$, the relative phase angle $\phi$, the $X$ component $V_{2\omega}^X = V_{2\omega}^R \cos\phi$ and the $Y$ component $V_{2\omega}^Y = V_{2\omega}^R \sin\phi$. Here $R$ and $\phi$ provide a complete description of $V_{2\omega}$; similarly, $X$ and $Y$ components provide a complete description of $V_{2\omega}$. Since the amplitude $V_{2\omega}^R$ is always positive, we combine the sign of $\sin\phi$ and $V_{2\omega}^R$, and define the $V_{2\omega} = sign(sin\phi) \cdot V_{2\omega}^R = sign(sin\phi)|V_{2\omega}|$ to showcase the $cos\theta$ dependence of the SHST signal. All $\theta$ dependent $V_{2\omega}$ data shown in the paper are

original data without subtracting any offset signal. The temperature of the measurement ranges from 2K to 300K. The applied magnetic field is parallel to our sample plane and the maximum field is 14T.

**Nonlocal SHST signal $V_{2\omega}$ vs. B and $I_{in}$**

Supplementary Information Fig. S4a displays the corresponding SHST signal $V_{2\omega} = sign(sin\phi)|V_{2\omega}|$ of which shown in Figs. 2a-b in the main text. Two regions can be observed in Supplementary Information Fig. S4a, one with positive $V_{2\omega}$ signal and the other with negative $V_{2\omega}$ signal. These two types of signals are further illustrated in Supplementary Information Figs. S4b-c and Supplementary Information Figs. S4d-e, respectively. Supplementary Information Figs. S4b and S4c plot $V_{2\omega}$ and the corresponding phase $\phi$ as a function of magnetic field angle $\theta$, for which the data was collected with an AC injection current with root mean square value of $I_{in} = 100\mu A$ and at $B = 14T$. Due to the orthogonal rule of ISHE at the detector, $V_{2\omega}$ has a $cos\theta$ dependence with a positive value at $\theta = 0$. The corresponding phases $\phi$ are $\pm 90°$. The features shown in Supplementary Information Figs. S4b-c are consistent with thermally generated magnons in magnetically order materials [5]. Whereas for $B = 3T$ with the same injection current, $V_{2\omega}$ has a $cos\theta$ dependence with an opposite sign, and the corresponding phases $\phi$ are around -50° and 130°, deviating from ±90°, as shown in Supplementary Information Figs. S4d-e. Such deviation of $\phi$ from ±90° indicates a nonzero $V_{2\omega}^X$, which is one major characteristics of the anomalous spin transport in GGG.

**$V_{2\omega}^X$ and $V_{2\omega}^Y$ vs. T and B at different channel lengths**

In order to understand how spin angular momenta are transported in GGG, we studied the temperature and magnetic field dependence of the nonlocal SHST signal $V_{2\omega}^X$ and $V_{2\omega}^Y$ at different injector-detector distances. Supplementary Information Figs. S7a-b replot the $V_{2\omega}^X$ and $V_{2\omega}^Y$ for $d = 1~\mu m$ in Figs. 3a-b in the main text, and the inset shows the temperature dependent evolution of $V_{2\omega}^X$ and $V_{2\omega}^Y$ under magnetic field of 2.1 T. Similar to Supplementary Information Figs. S7a-b, Supplementary Information Figs. S7c-d and Supplementary Information Figs. 7e-f are showing the phase diagrams of $V_{2\omega}^X$ and $V_{2\omega}^Y$ vs. $T$ and $B$ for channel length $d = 2~\mu m$ and $d = 4~\mu m$, respectively. Comparing these phase diagrams of different distances, we find that the normal SHST signal (red parts in Supplementary Information Figs. S7b, d, e) disappears quickly when the channel length becomes longer than 2μm. On the contrary, the anomalous SHST signal (blue parts in Supplementary Information Figs. S7a-f) decreases much slower, which means that spin angular momentum can transport to far greater distance in GGG in the anomalous state.

**Monte Carlo simulations**

We consider the following Hamiltonian:

$$\mathcal{H} = J_1 \sum_{\langle ij \rangle} \boldsymbol{S}_i \cdot \boldsymbol{S}_j + D r_{nn}^3 \sum_{j>i} \left[ \frac{\boldsymbol{S}_i \cdot \boldsymbol{S}_j}{|\boldsymbol{r}_{ij}|^3} - \frac{3(\boldsymbol{S}_i \cdot \boldsymbol{r}_{ij})(\boldsymbol{S}_j \cdot \boldsymbol{r}_{ij})}{|\boldsymbol{r}_{ij}|^5} \right] - g\mu_B \sum_i \boldsymbol{B} \cdot \boldsymbol{S}_i, \quad (1)$$

where $J_1$ is the super-exchange for nearest-neighbor bonds, and $D$ is the strength of the dipole-dipole interaction. It is worth noting that the Hamiltonian, even in absence of dipole-dipole interaction, contains exotic "octupolar order" instead of conventional "dipolar order" at the lowest available temperature[50]. The magnitude of the classical spins $\boldsymbol{S}_i$ is fixed as $|\boldsymbol{S}_i| = 7/2$ in the simulation. We fix the $g$-factor $g = 2$ for $Gd^{3+}$. The value of the dipole-dipole interaction is estimated as $D \approx 0.0458K$ for a lattice constant of 1.2376nm. In the simulations, the dipole-dipole interaction was treated using the standard Ewald summation technique. The optimal value of $J_1 \approx 0.138K$ was extracted by fitting the specific heat data of GGG in Ref. [35] while fixing $D = 0.0458K$. Since we restrict to $T \geq 1$ K in the simulation, the relatively small further-neighbor super-exchanges can be safely neglected [51].

We employed the classical Monte Carlo (MC) simulations for the model on a $4 \times 4 \times 4$ lattice (each unit cell contains 24 sublattices) with periodic boundary conditions. Specifically, a total of $10^5$ MC sweeps with Metropolis update were used for each parameter set. During the first $2.5 \times 10^4$ sweeps the system was annealed from $T = 15$ K to the target temperature, and the thermodynamic properties were measured using the $5 \times 10^4$ sweeps at the end of the calculation. For each parameter set a total of 8 independent MC runs (with different random seeds) were performed to calculate the statistical averages and error bars. We have verified that the Metropolis update scheme is sufficient for $T \geq 1$ K and there is no obvious system size dependence comparing the results on $4 \times 4 \times 4$ and $8 \times 8 \times 8$ lattices (see datails in Supplementary Information Fig. S15).

The simulated $B$-$T$ phase diagrams of magnetic specific heat $C$, magnetization $M$ are shown in Supplementary Information Fig. S8 and the magnetic susceptibility $\chi$ is shown in Fig. 4c in the main text. From the susceptibility data, it's obvious that spin fluctuations become stronger below 5 K and below the saturation field, suggesting the experimentally observed anomalous spin transport is rooted in the spin fluctuations and spin-spin correlations. Moreover, the relevant experimental parameter regime in the finite $V_{2\omega}^X$ distribution shown in Fig. 3a overlaps with the critical fan of the quantum critical point at saturation.

The spin waves in the paramagnetic phase were computed by the equation-of-motion (EOM) method using spin configurations sampled from Monte Carlo simulations on $4 \times 4 \times 4$ systems. Specifically, we integrated the Landau-Lifshitz equation in the 4th-order Adams predictor-corrector scheme, with a total duration of 20000 K$^{-1}$ and step size $\Delta t = 0.002$ K. Supplementary Information Fig. S13 shows the dynamic spin structure factor $S(k, w)$ at 2K and under different magnetic fields. At lower magnetic field, while there is no long-range magnetic order, we can still observe diffusive spin-spin correlation; with increasing magnetic field, the magnon bands are increasingly well-defined. In Supplementary Information Fig. S14, the magnon life time $\tau = \frac{1}{\sigma}$ can be obtained by a Gaussian fit $A \cdot e^{-\frac{(\omega-\mu)^2}{2\sigma^2}}$ to the energy dispersion, where $\mu$ is the band center and $\sigma$ is the standard deviation representing linewidth. We plot the magnon life time of the highest band at $\Gamma$ point as a function of magnetic field as shown in Fig. 4d in the main text. Note that for $B = 0$ and 0.71T, it is difficult to get a reliable fitting curve, meaning there is no well-defined magnon, so that they are not included. For $B = 1.41$T to 5.66T, reliable and rapidly increasing magnon life time can be obtained, which is shown in Fig. 4d. Importantly, the experimentally observed dominance of normal spin transport with drastic decrease in spin transport length is connected to the formation of conventional gapped magnon bands under strong external magnetic field in the Monte Carlo simulations.

**Distance dependence of normal spin transport in GGG and MnPS$_3$**

We investigate two points in the "normal" area of the *B-T* phase diagram, one is at the low temperature and high magnetic field regime ($T = 2$K, $B = 14$T, $I_{in} = 100\mu A$), and the other one is at the high temperature and low magnetic field regime ($T = 5$K, $B = 3$T, $I_{in} = 100\mu A$). The experimental data were fitted to the following equation derived from a two-dimensional diffusion model [11]:

$$V_{2\omega}(d) = CK_0(d/\lambda) \tag{2}$$

where $K_0(d/\lambda)$ is the modified Bessel function of the second kind, $\lambda$ is the spin diffusion length, and $C$ is a numerical coefficient that does not depend on distance $d$. Here the two-dimensional diffusion model [11] is used because the thickness of GGG is 500 $\mu m > d$. A typical distance dependent $V_{2\omega}$ in MnPS3 is also plotted in Supplementary Information Fig. S9. For MnPS3, the one-dimensional magnon diffusion model [5] is used since its thickness is around 30nm $\ll d$. These normal SHST signals almost fade away within a distance of 5 $\mu m$.

**Finite element analysis of temperature gradience vs. distance**

The signal decay curve for spin transport distance from $4\mu m$ to $100\mu m$ is compared with the temperature gradient distribution obtained from a finite element thermal analysis in Supplementary Information Fig. S11. The anomalous $|V_{2\omega}|$ decays much slower than the temperature gradient, indicating that the signal cannot be resulted from a local spin Seebeck effect driven by local temperature gradience beneath the detecting electrode.

In the finite element simulation, $t_{GGG}$ = 500μm is the GGG thickness, and both the width and length of the GGG slab are 5mm. The injector has a thickness of $t_{Pt}$ = 10nm, a length of $l_{Pt}$ = 100μm and a width of $w_{Pt}$ = 200nm. The resistance of the Pt injector is measured to be 7000Ω with an injection current of $I_{in} = 100\mu A$. The heat current normal to the GGG|vacuum and Pt|vacuum vanish. At the bottom of the GGG slab, the boundary condition $T$ = 2K is used. Other parameters used in the finite element analysis are listed in Supplementary Information Table 1.

**Separation of normal and anomalous spin transport signals**

It is experimentally observed that 1) normal $V_{2\omega}^X = 0$ and 2) normal $V_{2\omega}^Y$ decays quickly. These two points can be exploited to separate the normal contribution and the anomalous contribution of the spin transport signal to $V_{2\omega}^X$ and $V_{2\omega}^Y$ when they are mixed at lower temperature and lower magnetic field. Supplementary Information Fig. S12a plots $V_{2\omega}^X$ and $V_{2\omega}^Y$ separately of the $|V_{2\omega}|$ data shown in Supplementary Information Fig. S11. One can see that $V_{2\omega}^X$ follows one monotonic curve while $V_{2\omega}^Y$ is non-monotonic, highlighting the fact that $V_{2\omega}^X$ only has anomalous contribution while $V_{2\omega}^Y$ has both anomalous and normal contributions. A fitting of $V_{2\omega}^X$ with the two-dimensional diffusion model gives $\lambda = 55.276 \pm 6.53\mu m$, which is one order of magnitude bigger than the normal case. The isolation of the normal spin signal can be based on the observation that the normal spin transport length is relatively short and the trends of $V_{2\omega}^X$ vs. $d$ and $V_{2\omega}^Y$ vs. $d$ are similar for distance over $20\mu m$. With the above information, we raise the ansatz that the anomalous SHST signal $^AV_{2\omega}^Y$ is proportional to $V_{2\omega}^X$. Thus, the normal SHST signal $^NV_{2\omega}$ can be extracted from $V_{2\omega}^Y$ by subtracting $V_{2\omega}^Y$ with the anomalous SHST signal $^AV_{2\omega}^Y = -2.58 \times V_{2\omega}^X$ (Supplementary Information Fig. S12b). Here the ratio -2.58 is obtained by setting $^{N}V_{2\omega} = 0$ at $d = 60\ \mu m$. A fitting of the extracted $^{N}V_{2\omega}$ data (blue points in Supplementary Information Fig. 12b) to equation (2) gives $\lambda = 7.3999 \pm 1.79 \mu m$, which is consistent with the normal spin transport behaviors shown in Supplementary Information Fig. S9.

## Data Availability

The experimental data generated in this study (Fig. 1-5 in the main text and Fig. S1-7, Fig. S9-12, Fig. S16-17 in the Supplementary Information) have been deposited in the Harvard Dataverse under accession code *CL317* [52]. All other data that support the findings of this study are available from the corresponding authors upon request.

## Code Availability

The code supporting the analysis reported in this paper is available from the corresponding authors upon request.

## Acknowledgments

This project has been supported by the National Key R&D Program of China (Grant Nos. 2024YFA1409001 (J.-H.C), 2024YFA1408303 (Z.W.)), the Innovation Program for Quantum Science and Technology (2021ZD0302403 (J.-H.C)), the National Natural Science Foundation of

China (NSFC Grant Nos. 12304537 (D.C.), 92265106 (J.-H.C), 11774010 (J.-H.C), 11921005 (J.-H.C), 12374124 (Z.W.)). J.-H.C acknowledges technical support form Peking Nanofab.

## Author Contributions Statement

J.-H.C conceived the idea and directed the experiment; D.C. fabricated most of the devices and performed the transport measurements; B.L. & L.J. aided in device fabrication; C.T. did the SQUID measurement; S.Q., G.C., S.Y., M.H., J.C. & Y.F. aided in transport measurement; D.C., B.L. & K.C. collected and did data analysis; J.Z. provided computational facilities; L.X., J.Z., Z.X., X.C.X., R.S., Z.W. & J.X. provided theoretical and numerical analysis; H.C. performed the finite element analysis of heat distribution; D.C. & J.-H.C wrote the manuscript; all authors commented and modified the manuscript.

## Competing interests statement

The authors declare no competing interests.

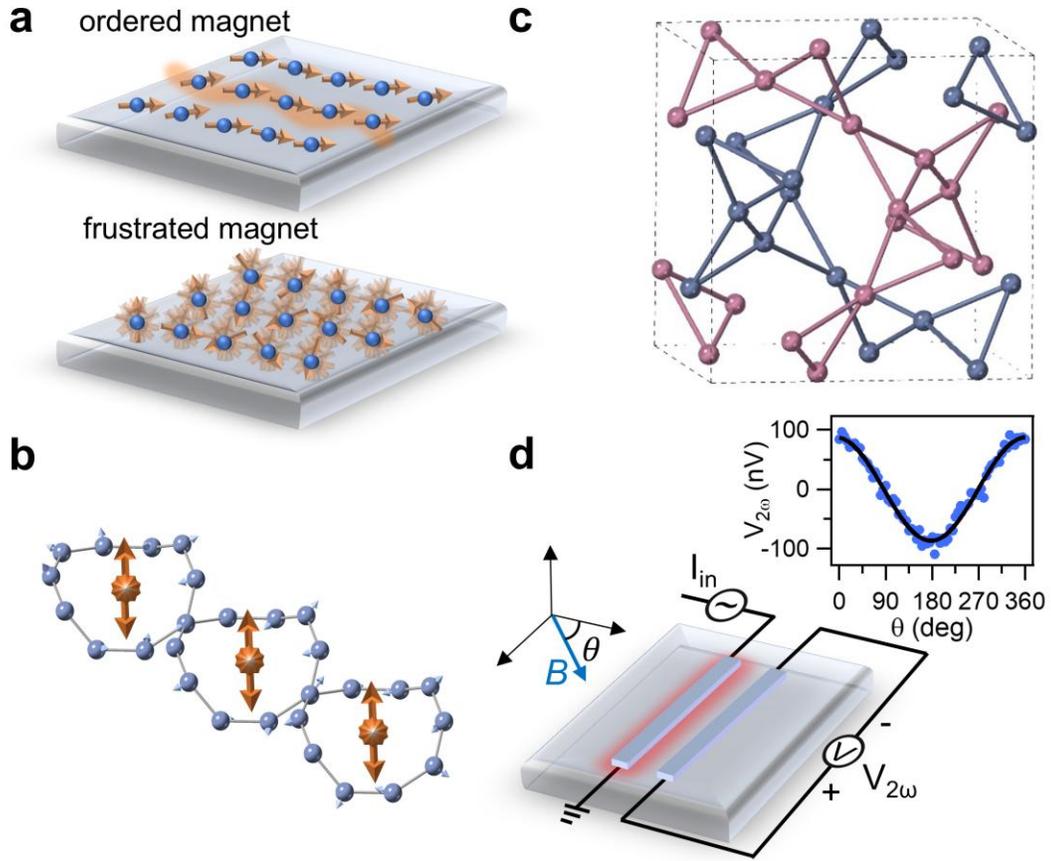

**Fig. 1 | Spin transport in frustrated hyperkagome magnet GGG.** (**a**) Sketches of spin wave in ordered magnet and fluctuating spins in frustrated magnet. The opaque arrows represent the localized magnetic moments, while the transparent arrows represent the fluctuations of the moments. (**b**) Schematic of the long-range multipole order formed from ten-spin loops of $Gd^{3+}$ in GGG. Orange double-headed arrows indicate the ten-spin director that are ordered normal to the loop plane. Smaller gray arrows represent the localized magnetic moments. (**c**) Crystal structure of hyperkagome magnet GGG with only Gd atoms shown. The two inter-penetrating corner-sharing $Gd^{3+}$ triangular sublattices are colored in gray and pink, respectively. (**d**) Schematics of the nonlocal spin transport measurement setup with electrical connections. Two Pt wires are patterned on a GGG slab, one to inject spin current by thermal excitation and the other to detect the spin current through inverse spin Hall effect. Specifically, $I_{in}$: AC injection current; $V_{2\omega}$: the second harmonic thermal magnon inverse spin Hall signal; θ: the angle of the in-plane magnetic field **B** with respect to the direction perpendicular to the Pt electrodes. The inset shows angle dependent $V_{2\omega}$ of GGG with $B = 3T$ and $I_{in} = 80\mu A$ and $T= 2K$, in which the black solid line is a cosine function fit.

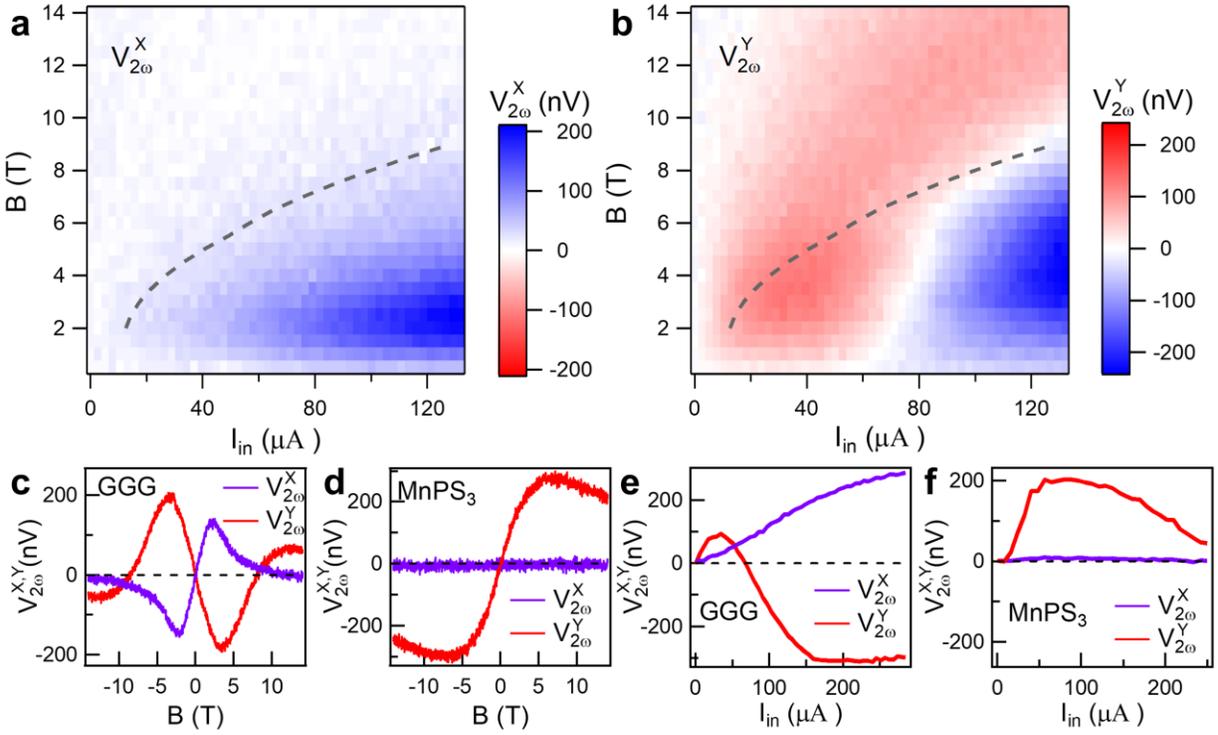

**Fig. 2 | Magnetic field and injection current dependent nonlocal signal $V_{2\omega}^{X,Y}$.** (a) (b) 2D color plot of $V_{2\omega}^{X}$ (a) and $V_{2\omega}^{Y}$ (b) versus $B$ and $I_{in}$ in GGG, respectively. All data are taken from a GGG device with a $2\mu m$ channel length at $T = 2K$. The dashed gray lines are guides for the boundary of the anomalous signal. (c) (d) $V_{2\omega}^{X}$ and $V_{2\omega}^{Y}$ versus $B$ in a typical GGG (c) and MnPS$_3$ (d) device with $I_{in} = 100\mu A$, respectively. (e) (f) $V_{2\omega}^{X}$ and $V_{2\omega}^{Y}$ versus $I_{in}$ in a typical GGG device at $T = 2K$ and $B = 3T$ (e) and a typical MnPS$_3$ device at $T = 2K$ and $B = 4T$ (f), respectively. The injection current frequency is 17.77Hz.

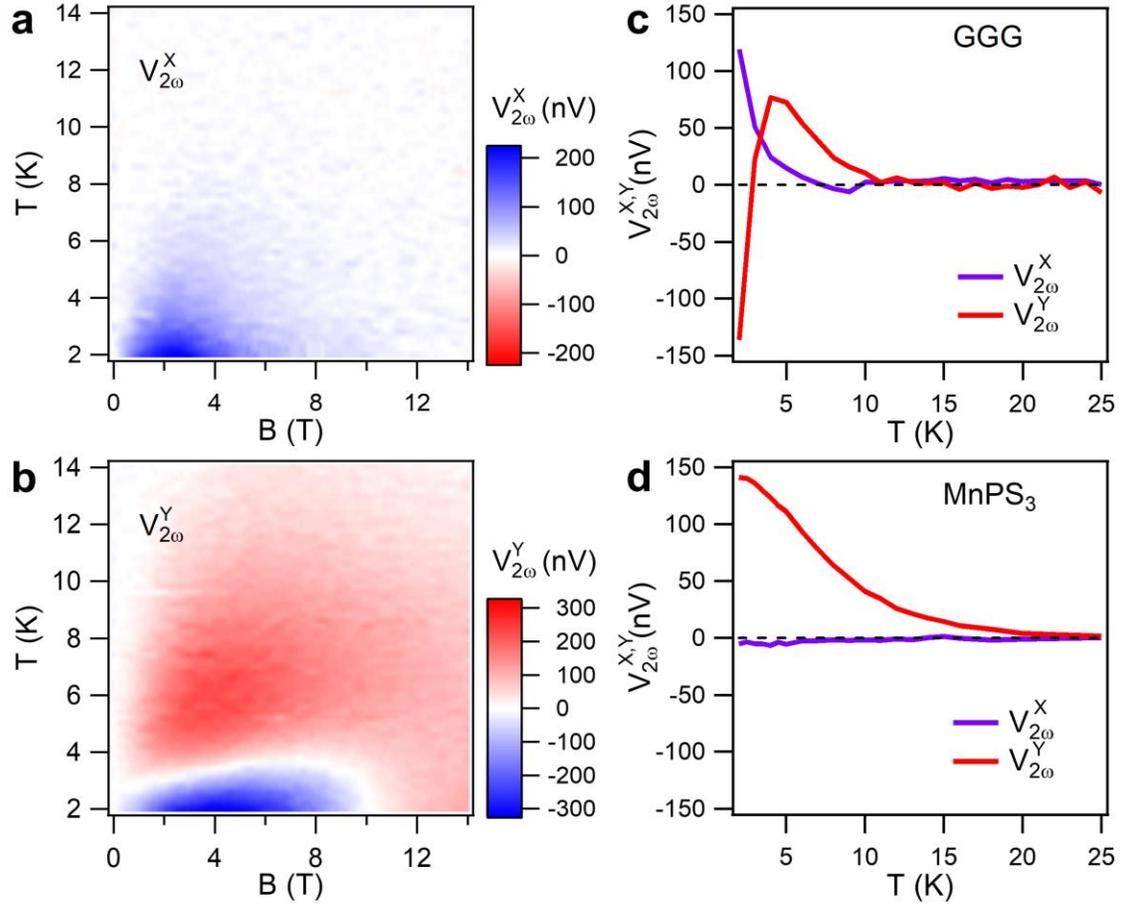

**Fig. 3 | Temperature and magnetic field dependence of the nonlocal signal $V_{2\omega}^{X,Y}$.** (a) (b) 2D color plots of $V_{2\omega}^{X}$ and $V_{2\omega}^{Y}$ vs. $T$ and $B$, respectively, with $I_{in} = 100\mu A$ and channel length of $1\mu m$. (c) (d) $T$ dependence of $V_{2\omega}^{X}$ and $V_{2\omega}^{Y}$ in a typical GGG device (c) and MnPS$_3$ device (d), respectively. Here $B = $ 3T for GGG and 4T for MnPS$_3$, and the injection current frequency is 17.77Hz.

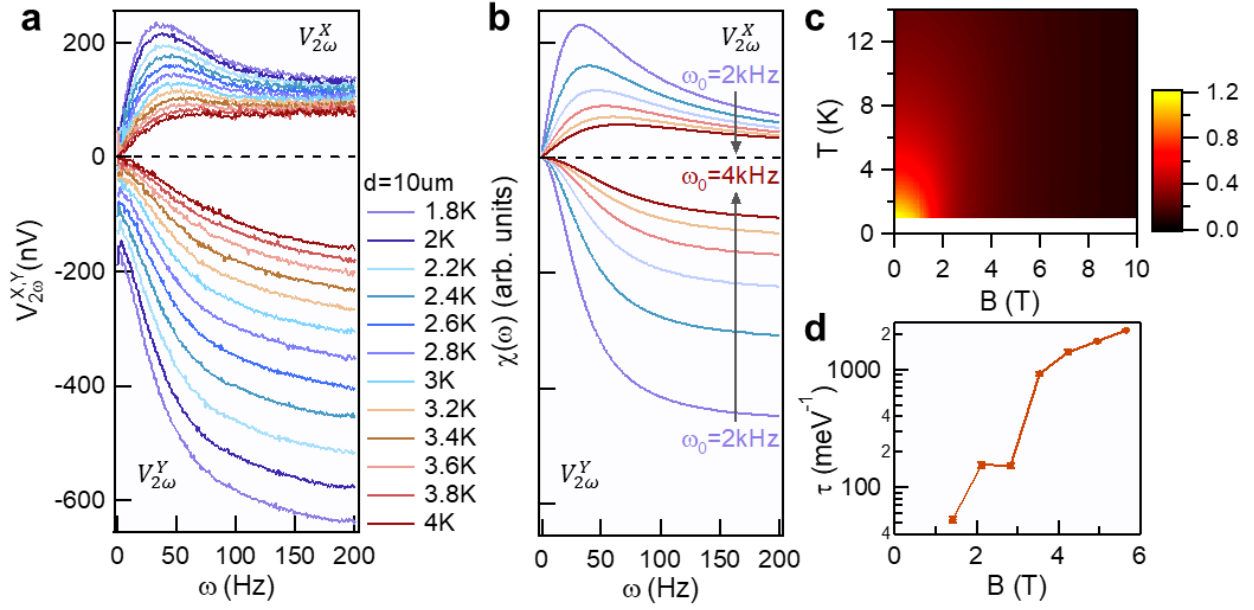

**Fig. 4 | Magnetic excitation related to spin-spin correlations. (a)** Injection frequency-dependent $V_{2\omega}^X$ and $V_{2\omega}^Y$ at 1.8K to 4K with $I_{in} = 200\mu A$, $B = 3T$ and channel length of $10\mu m$. **(b)** Simulation of the real and imaginary part of the response $\chi(\omega)$ for an overdamped forced oscillator. Here $\omega_0 = 2, 2.4, 2.8, 3.2, 3.6, 4\ kHz$ and $\beta = 60$. **(c)** Monte Carlo simulated magnetic susceptibility vs. $T$ and $B$ applied along the [-1,1,0] direction. **(d)** Magnetic field dependent magnon life time $\tau$ of the highest band at $\Gamma$ point obtained from Monte Carlo simulations at 2K. The error bars of $\tau$ are calculated from the standard error of the linewidth $\sigma$.

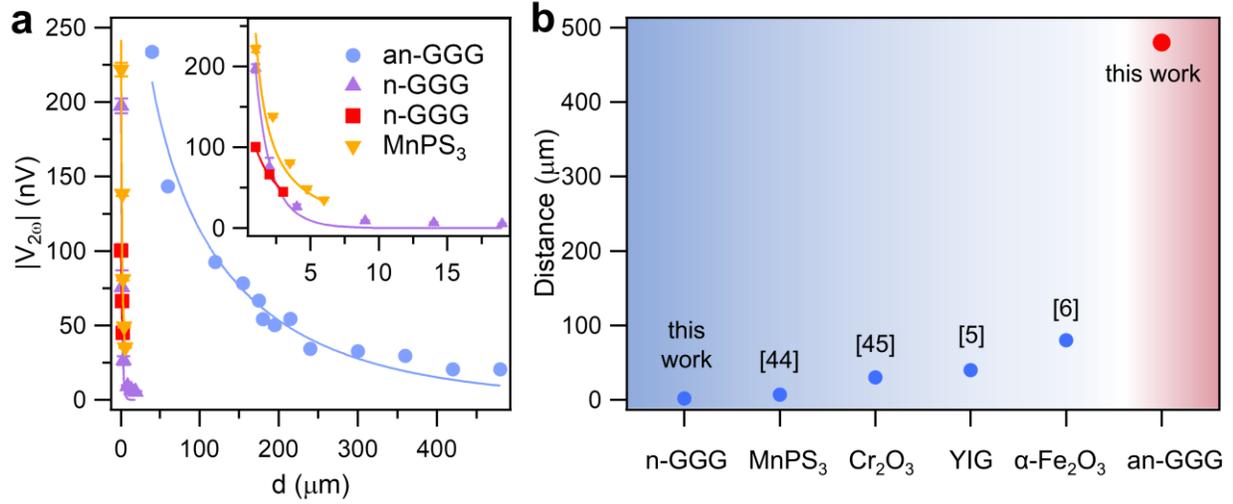

**Fig. 5 | Anomalous long-distance spin transport in GGG.** (**a**) $|V_{2\omega}|$ versus channel length $d$ for normal and anomalous spin transport states. Anomalous GGG (marked as an-GGG): blue circles, $B = 3$T, $I_{in} = 200\mu A$, $T = 2$K; normal GGG (marked as n-GGG): purple triangles, $B = 3$T, $I_{in} = 100\mu A$, $T = 5$K; red squares, $B = 14$T, $I_{in} = 100\mu A$, $T = 2$K; MnPS$_3$: yellow triangles, $B = 4$T, $I_{in} = 100\mu A$, $T = 2$K. Every data point is extracted by a cosine fit from a magnetic field angle-dependent measurement, with error bar represented the standard error of fit. The yellow solid line is a fit to the 1D magnon diffusion model [5] while other solid lines are fits to the 2D diffusion model due to larger sample thickness [11]. (**b**) Comparison of our work and the reported maximum transport distances of thermal excited magnons in the literature [5, 6, 44, 45], grouped in magnetically ordered region (blue part), and magnetically frustrated region (red part).